\pdfoutput=1
\documentclass[aps,pre, twocolumn, groupedaddress]{revtex4-1}

\usepackage{lipsum}
\usepackage{float}
\usepackage{mathtools}
\usepackage{graphicx}
\usepackage{subfigure}
\usepackage{caption}
\usepackage{dcolumn}
\usepackage{amsmath}    
\usepackage{amssymb}
\usepackage{bm}
\usepackage{hyperref}
\usepackage{latexsym}
\usepackage{verbatim}
\usepackage[normalem]{ulem}
\usepackage{color}
\usepackage{parskip}
\def\beq{\begin{equation}}
\def\eeq{\end{equation}}

\def\beq{\begin{equation}}                          
\def\eeq{\end{equation}}                          
\def\bea{\begin{eqnarray}}                          
\def\eea{\end{eqnarray}}

\DeclareRobustCommand{\uvec}[1]{{%
  \ifcsname uvec#1\endcsname
     \csname uvec#1\endcsname
   \else
    \bm{\hat{\mathbf{#1}}}%
   \fi
}}
\preprint{}
\begin{document}
\preprint{}
\title{ Current reversal in polar flock at order-disorder interface}
\author{Jay Prakash Singh$^{1,2}$}
\email{jayps.rs.phy16@itbhu.ac.in}

 \author{Partha Sarathi Mondal$^{1}$}
 \email{parthasarathimondal.rs.phy21@itbhu.ac.in}
 \author{Vivek Semwal$^{1}$}
 \email{viveksemwal.rs.phy17@itbhu.ac.in}

 \author{Shradha Mishra$^{1}$}
 \email[]{smishra.phy@itbhu.ac.in}

\affiliation{
$^{1}$ Indian Institute of Technology (BHU) Varanasi, India 221005\\
$^{2}$ Israel Institute of Technology,Technion Hafia Israel\\
}

\date{\today}

\begin{abstract}
    {	We   studied a system of polar self-propelled particles (SPPs)
	on a thin rectangular channel designed into three regions of order-disorder-order. 
The division of the three regions is made on the basis of the noise SPPs experience in the respective regions. The noise in the two wide region is chosen lower than the critical noise of order-disorder transition and noise in the middle region or interface
	is higher than the critical noise. This make  the geometry  of the system analogous to the Josephson Junction (JJ) in solid state physics. 
	Keeping all other parameters fixed, we study the properties of the moving SPPs in the bulk as well as along the interface for different widths of the junction. 
 On increasing  interface width, system shows a order-to-disorder transition 
	from coherent moving SPPs in the whole system to the interrupted current for large interface width. 
	Surprisingly, inside the interface we observed the current reversal for intermediate widths of the interface. Such current reversal is due to the strong randomness present inside the interface, that makes  the  wall of the interface reflecting. 
 Hence 
	Our study  give a new interesting  collective properties of SPPs at the interface 
	which can be useful to design  devices like switch using active agents.  }
    \end{abstract} 
    
 \maketitle
 
 \section{Introduction \label{Introduction}}
Emergence of collective motion \cite{Nedelec1997,Yokota1986,Toyo,kron,Laub} and global ordering \cite{Sudipta,VicsekT,Bond1,Mis1,TonerTu1995,Singh1} among the various living or non-living systems, are well known phenomena. Each particle  show  systematic motion at the cost of its internal energy. All individuals in a group synchronize themselves to show different behavioral state, exhibiting host of interesting  properties like pattern formation \cite{Pattern}, non equilibrium phase transition \cite{VicsekT}, large density fluctuation \cite{Pattern,Greg,Chate2008,Bhatta,Narayan}, enhance dynamics \cite{Bechi1,Angel1,Hardy1,Skumar1,Pablo1,Vivek1,Patta2019,Viv1,Bhask1,Jay1}, motility induced phase separation \cite{Mips1,Mips2,Mips3,Mips4,Mips5} etc,. Interestingly, different real biological systems are encountered with different kinds of confined geometry \cite{alberts2002,Alb1,Das1,Das2,Das3,Das4}. Confinement and boundary play significant role in variety of biological systems \cite{Alb1}, sheared systems \cite{Sara2}, other places like in fluid dynamics. Boundary can induce many interesting phenomena like, spontaneous flow inside the channel \cite{RB}, and another classic example include Rayleigh-Benard convection in the fluid \cite{Bod3}. There are variety of practical applications based on confined geometry like; mass transport in nanofluids to enhance the microfluidic devices \cite{Zum4,voi2005}, geophysical applications etc.,\\
There are few studies where researchers have seen the behaviour of SPPs at the interface of two different substrate media. Most of studies involves the media of two dissimilar fluids \cite{Int1,Int2,Int3,Int4}. For example, Dirichlet et al. \cite{Kilian} observed that catalytic active Brownian microswimmer at different solid-liquid interface shows inhomogeneity in the particles speeds with respect to the orientation of catalytic substrate at different interface. Another well known classical example of interface between superconductor and insulator with boundary is Josephson Junction (JJ) in solid state system \cite{currentR,JJ1,JJ2, JJ3, JJ4}. 

Motivated with the JJ in solid state, here in this article we  will discuss the collective properties of SPPs by designing a setup analogous to JJ. We have modeled a system of polar SPPs with alignment interaction through a thin rectangular narrow channel. Further, the thin channel is divided into three regions wherein two opposite regions, SPPs move coherently. In the middle region SPPs diffuse randomly and no net current. 
Although the comparison between our setup and Josephson junction is not very common, since the  superconducting phenomena are macroscopically quantum in nature, but we still designed an analogous model system for collection of SPPs and observe the properties of it.
\\
We also studied the case where  a small external field is introduced,  along the long axis of the system which gives an easy direction for particle alignment. System is studied for different widths of the intermediate disorder region with and without easy direction (perturbation). On tuning the width of the disorder region, SPPs with perturbation shows a non-equilibrium phase transition whereas without perturbation, it shows a weak dependence on the interface width. Further, at the junction, particles get reflected from it's walls and we  observed the
current reversal. This lead to the 
some of the SPPs to move in opposite direction and hence contribute in negative current. We also showed one application of
the system geometry to use it for sorting two different type of SPPs.\\

Rest of the manuscript is divided  as follows; we have discussed the details of the model  in section \ref{model}. The results are discussed in section \ref{results}, one possible application in section \ref{secsort} and conclusion of the paper with summary is discussed in section \ref{summary}.\\
\begin{figure}[ht]
\centering
\includegraphics[width=1.0 \linewidth]{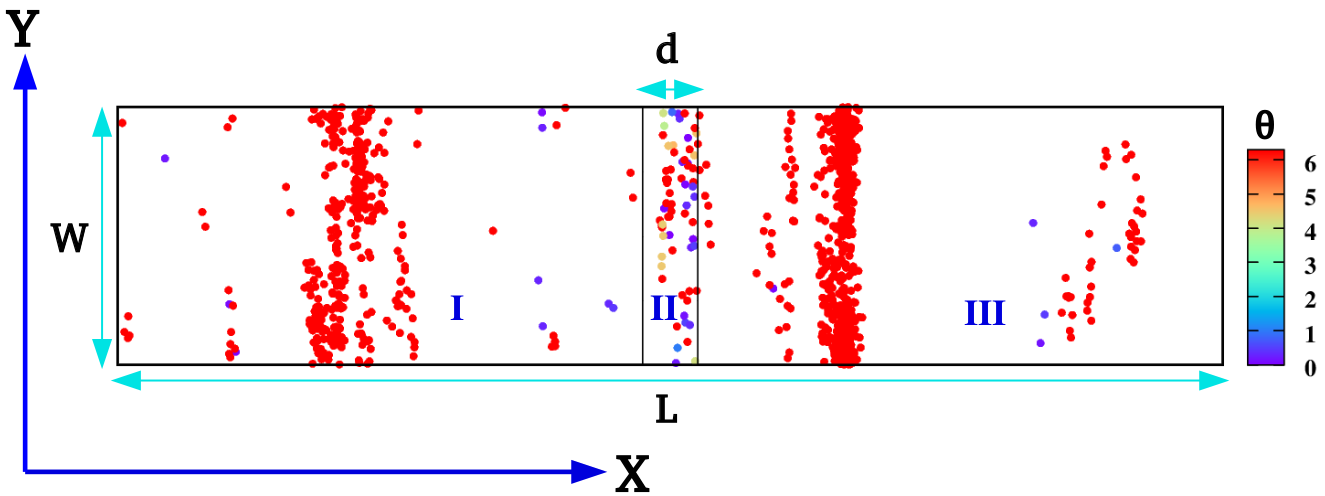}
\caption{(color online) We show a model picture of the system obtained from the simulation in which different color  shows particle's  orientation $\theta$ ($\in [0,2\pi]$). The region II  is the interface (junction) with disorder  region where as  the region to the left and right of the interface (I $\&$ III) is the ordered region. Noise strength in region II is $\eta_{II}=0.7$ and in regions I, III $\eta_{I}=\eta_{III}=0.3$. $d$ is the width of the interface.The two vertical lines in the middle indicate the boundary of the interface.
 The $W$ and $L$ are the short and long dimensions of the system. The axis for the two directions of the system is  drawn on the left. Periodic boundary condition is used in both the directions.} 
\label{fig:fig1}
\end{figure}

\section{Model \label{model}}
We consider a collection of polar self-propelled particles (SPPs) moving on a two-dimensional
substrate on a rectangular narrow channel with periodic boundary conditions (PBC) in both directions. The short and 
long axis of the channel are denoted by $W$ and $L$, respectively as shown in fig.{\ref{fig:fig1}}. Particles  interact through a short-range alignment interaction within a small interaction
radius $R_0 = 1$\cite{VicsekT}. 
Moreover, the strength of interaction of each SPP is the same.
The system is partitioned  into three regions: the two regions on the left and right  represent 
the ordered region  and third middle section shows the disordered region.
The middle disordered region is termed as junction or interface and the width $d$ of the interface is our 
tuning parameter. 
The width of the junction is varied from $d$ $ =(1-30)$. In three regions, 
each particle is defined by its position 
${\bf r}_i(t)$ and orientation $\theta_{i}(t)$ at time $t$ and they move along 
the direction of their orientation with a fixed speed $v_0=0.5$. 
The position and orientation updates of a particle are given by:
\begin{equation}
{\bf {r}}_{i} (t + \Delta{t}) = {\bf {r}}_i(t) + {v_{0}}{\bf n}_{i}{\Delta{t}}    
\label{eq1}
\end{equation}
\begin{equation}
	{\bf n}_{i}(t+\Delta{t})=\frac {{}\sum _{{j\in R_{0}}}{\bf n}_{j}(t)+\eta_{i,k} N_{i}(t){\bf {\bf{{\xi}}}}_{i}(t)}{w_i{(t)}}
\label{eq2}
\end{equation}
where, $\Delta t=1.0$ is the unit-time step and ${\bf n}_{i}=(\cos\theta_i,  \sin\theta_i)$ 
is the unit direction vector of the $i^{th}$ particle. In eq.\ref{eq2}, the first term in the right hand side represents the short-range alignment interaction 
inside the interaction radius ($R_0$) of the $i^{th}$ particle. 
The second term $\xi_i(t) = (\cos(\phi_i(t), \sin(\phi_i(t)))$ on the right hand side of eq.\ref{eq2} 
denotes the vector noise which measures the error made by
 the particle, following its neighbors. 
 $\phi_i$ is uniform random angle $\in (-\pi, \pi)$, $N_{i}(t)$ denote
the  number of neighbors within 
the interaction radius of the $i^{th}$ particle at time $t$. Further, $\eta_{i,k}$ $(k=I$, $II$ and $III$) ($\eta_{I} = \eta_{III} = 0.3$ and $\eta_{II}= 0.7$) shows the strength of the randomness present in the system for the three regions.  
We choose the mentioned values of noise, because for the clean polar SPPs interacting through Vicsek type alignment interaction  with vector noise, the order-disorder transition occurs at $\eta \sim 0.6$ (for the same set of parameters used here)\cite{jp,chate}. $w_{i}(t)$ is the normalisation factor which reduces the right hand side of the eq.\ref{eq2} to a unit vector. The above model we called as system without 
perturbation (WOP).\\
Due to the rectangular geometry of the system, particle experiences an easy axis for their motion (along the long axis of the system). We also introduced an external perturbation along the long-axis of the channel. It gives an easy direction for the SPPs motion; hence the orientation update equation will become 
\begin{equation}
{\bf n}_{i}(t+\Delta{t})=\frac {{}\sum _{{j\in R_{0}}}{\bf n}_{j}(t)+h_0{\bf {n_p}}+\eta_k N_{i}(t){\bf {\bf{{\xi}}}}_{i}(t)}{w_i{(t)}}
\label{eq3}
\end{equation}
here, $h_0$ is the strength of the external field and  kept fixed to a small value with direction ${\bf {n_p}}=(1, 0)$. Using the above equation eq.\ref{eq3}, the model  is referred to as system with perturbation (WP).
Further, number  density of SPPs is defined by $\rho=\frac{N}{L\times W }{=1.0}$, where $N$ is the total number of particles in the system. All the particles are allowed to move throughout the system and they experience the noise of different regions accordingly. We let the system evolve from random homogeneous state of density and orientation of particles. All the results discussed below are in the steady state, and total time step of the simulation is taken  ${10^6}$. One simulation step is counted after the update of all the particles once. Numerical details and  parameters are as chosen as $R_0=1.0$, $L=200, 400$, $W=5$, and $h_0$ is varied from $2\%$ to $6\%$  of the strength of alignment; which is fixed to $1$. A total 20  independent realisations are used for better statistics.

\section{Results\label{results}}
\subsection{Global ordering and  junction width ($d$)\label{Global orieantation order parameter}}
\begin{figure}[ht]
\centering
\includegraphics[width=1.0 \linewidth]{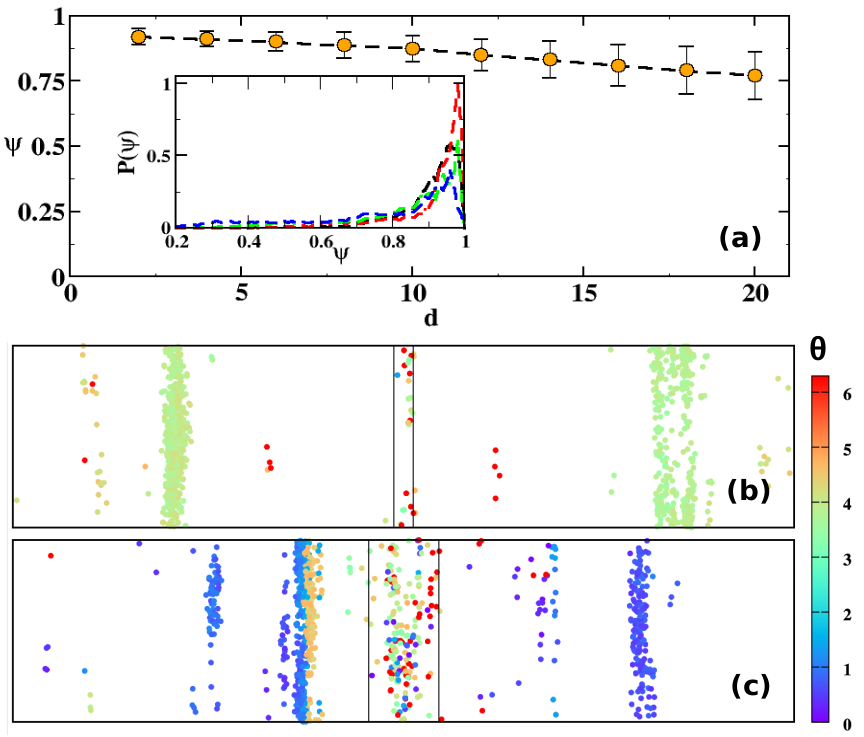}
\caption{(color online) All the plots (a)-(c) shown here for  without perturbation. (a) We plot the global orientation order parameter $\Psi$ vs. width of the disorder region $d$. Inset;(a) plot shows the global order parameter distribution $P(\Psi)$ for different width of the junction $d$. Different colored break lines are for $d$ = 4 (black), $d$ = 8 (red), $d$ = 12 (green) and $d$ = 18 (blue). Plot (b) and (c) show the space snapshots of the system for width $d$ = 5 and $d$ = 18 respectively. color of each particle represents it's orientation $\theta$ ($\in [0,2\pi]$) according to the color bar.  
$L\times{W}=200\times5$ and $\rho=1.0$} 
\label{fig:fig2}
\end{figure}

\begin{figure*}[ht]
\centering
\includegraphics[width=1.0 \linewidth]{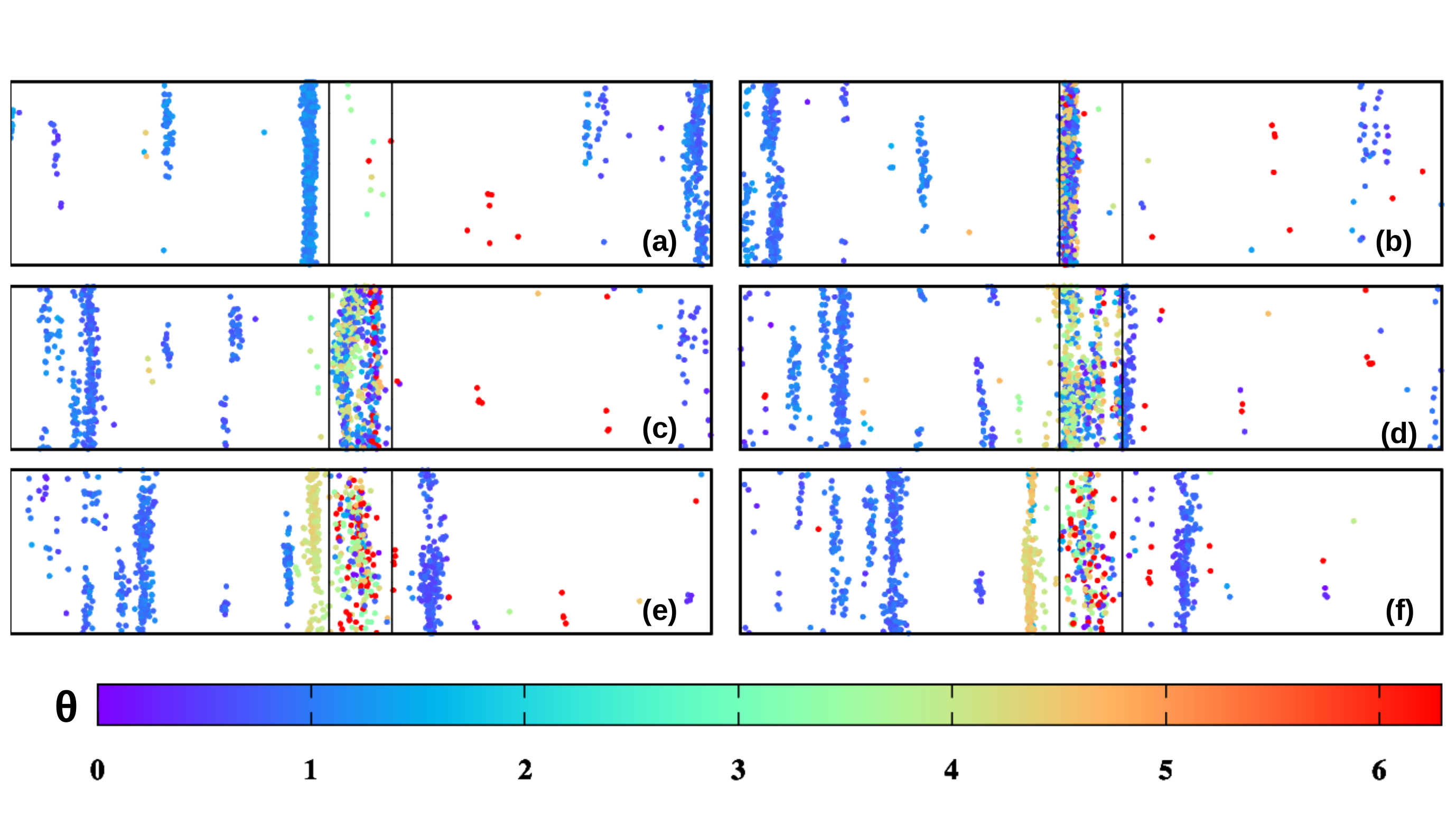}
\caption{
(color online) Snapshots of particles (for system WOP) with their orientation angle $\theta$ shown in color bar at times 
$t$ = 2122, 2163, 2204, 2223, 2253 and 2271 (a-f) respectively. The time is chosen  such that a dense band of particles moving towards the interface (blue color) in (a), enter inside the junction (b) and a part of the particles inside the band start getting reflected (having orientation towards $-x$ direction) (c-d) and macroscopic fraction of particles are reflected from the interface as shown in (e-f). Some part of particles are transmitted through the junction (e-f). 
The width of the interface $d=14$, $L\times{W}=200\times5$ and $\rho=1.0$}
\label{fig:fig3}
\end{figure*}

First, we study the effect of junction width $d$ on the global orientation in the whole system of size $L\times{W}=200\times5$ for different junction widths $d$. Ordering in the system is
characterized by the  orientation order parameter,
\begin{equation}
	\Psi(t)=\frac{1}{N}\vert{\sum^N_{i=1}}n_i(t)\vert	 
\label{eq4}
\end{equation}
In the ordered state, i.e., when majority  of particles are moving in the same direction, then $\Psi$ will be closer to $1$,
and of the order of $\frac{1}{\sqrt{N}}$ for a random disordered state. First, we show the variation of $\Psi = <\Psi(t)>$, where, $<..>$ means average over time in the steady state and over $20$ independent realisations. We first study the system WOP. In fig.\ref{fig:fig2}(a), we plot $\Psi$ {\em vs.}  junction width $d$ and found that with increase in $d$, $\Psi$ shows small decay, which have been further confirmed by orientation probability distribution function (PDF) $P(\Psi)$ in fig\ref{fig:fig2}(b). To understand  the small decay of $\Psi$ with  width $d$, we have shown the snapshots for two different junction widths $d=2$ and $18$ in fig.\ref{fig:fig2}(b) and (c), respectively.  The circles represent the particle and color of the circle shows their orientations. In general the SPPs form the ordered band inside the ordered region as shown by the dense moving SPPs along the channel in fig.\ref{fig:fig2}(b) (from right direction). The size of the ordered band or cluster
depends on the chosen set of system parameters \cite{shradhapre2010, Chate2008, ourdata}.  Hence for small width as shown in fig.\ref{fig:fig2}(b), when the width of the interface is smaller or of the order of the size of the band, the SPP passes the interface before it experiences the disorder present inside the interface. Hence the global order parameter retains high value $\sim 0.8$ and direction of moving SPPs band remains unaffected after interaction with the interface (shown by the almost a clear common orientation of all particles in the system in fig.\ref{fig:fig2}(b)). As we increase the width of the interface, and  the width of the disordered region is  larger than the size of the band. Then before the band can pass the interface it experiences the disturbance and front of the band randomise. Due to that the particles moving at the back also feel random orientation before they enter the interface. Hence, a part of the band reorient  before it can pass the interface as shown in fig.\ref{fig:fig3}(a-b). The whole process looks like a reflection due to the interface as in fig.\ref{fig:fig3}(c-d). Some  part of the band  able to come out of the interface from the other side and they contribute the forward moving current, but  a finite fraction of particles from the band observe reflection from the walls of the 
interface as shown in fig.\ref{fig:fig3}(e-f). A clear animation of the interaction of a band with the interface is shown in SM1. The interface acts like a partially reflecting wall, it leads the SPPs  to avoid the junction. Also to avoid the interface their orientation develops some contribution in $Y-$ direction as well. As shown in the snapshots fig.\ref{fig:fig3}(e-f), the orientation of particles are not strictly along $\theta = (0, \pi$ or $2 \pi)$. This reduces their frequency of going inside the interface. This results that mostly  the SPPs  are moving in the ordered region only, and we find  weak dependence of global order parameter  on the width $d$. 

\begin{figure}[ht]
\centering
\includegraphics[width=1.0 \linewidth]{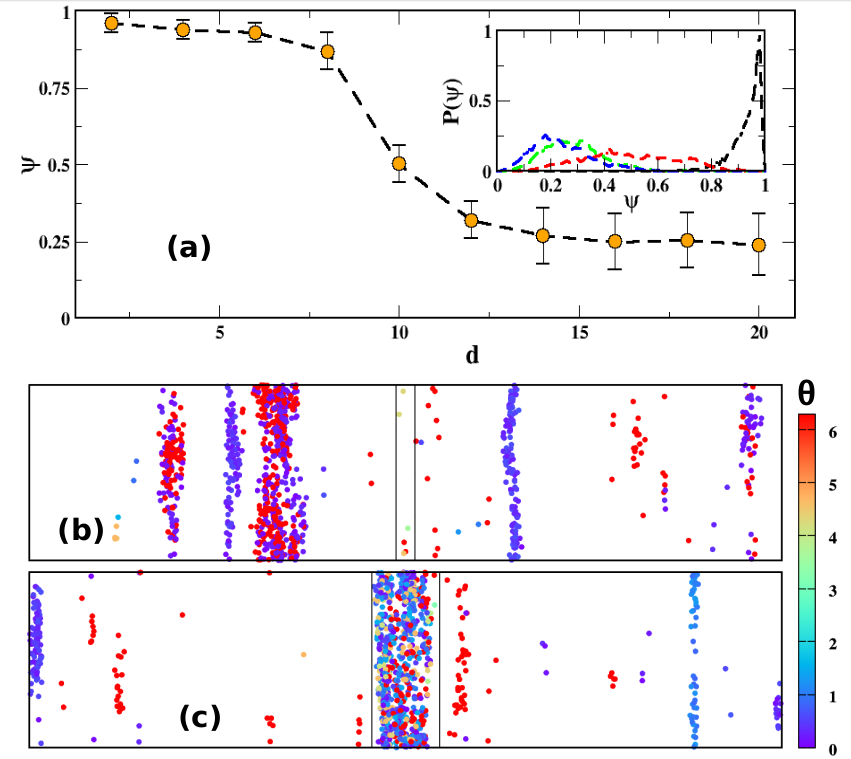}
\caption{(color online) All the plots (a)-(c) shown here for  with perturbation. (a) We plot the global orientation order parameter $\Psi$ vs. width of the disorder region $d$ with perturbation. Inset;(a) plot shows the global order parameter distribution $P(\Psi)$ for different width of the junction $d$. Different colored break lines are for $d$ = 4 (black), $d$ = 8 (red), $d$ = 12 (green) and $d$ = 18 (blue). color of each particle represents it's orientation $\theta$ ($\in [0,2\pi]$) according to the color bar. Plot (b) and (c) show the space snapshots of the system for width $d$ = 5 and $d$ = 18 respectively. The two vertical lines indicate the boundary of the interface.
$L\times{W}=200\times5$ and $\rho=1.0$}
\label{fig:fig4}
-\end{figure}

Further, we have studied the system WP. The perturbation introduced in such a way that flock is biased to move along the {\em +ve} direction of long axis ({\em x}-axis). Interestingly, we have found that global orientation order parameter $\Psi$ decay sharply with an increase in the junction width $d$ as shown in fig.\ref{fig:fig4}(a). Which has been confirmed by plotting the probability distribution function (PDF) $P(\Psi)$ for different junction widths $d$ in inset of  fig.\ref{fig:fig4}(a).

\begin{figure*}[ht]
\centering
\includegraphics[width=1.0 \linewidth]{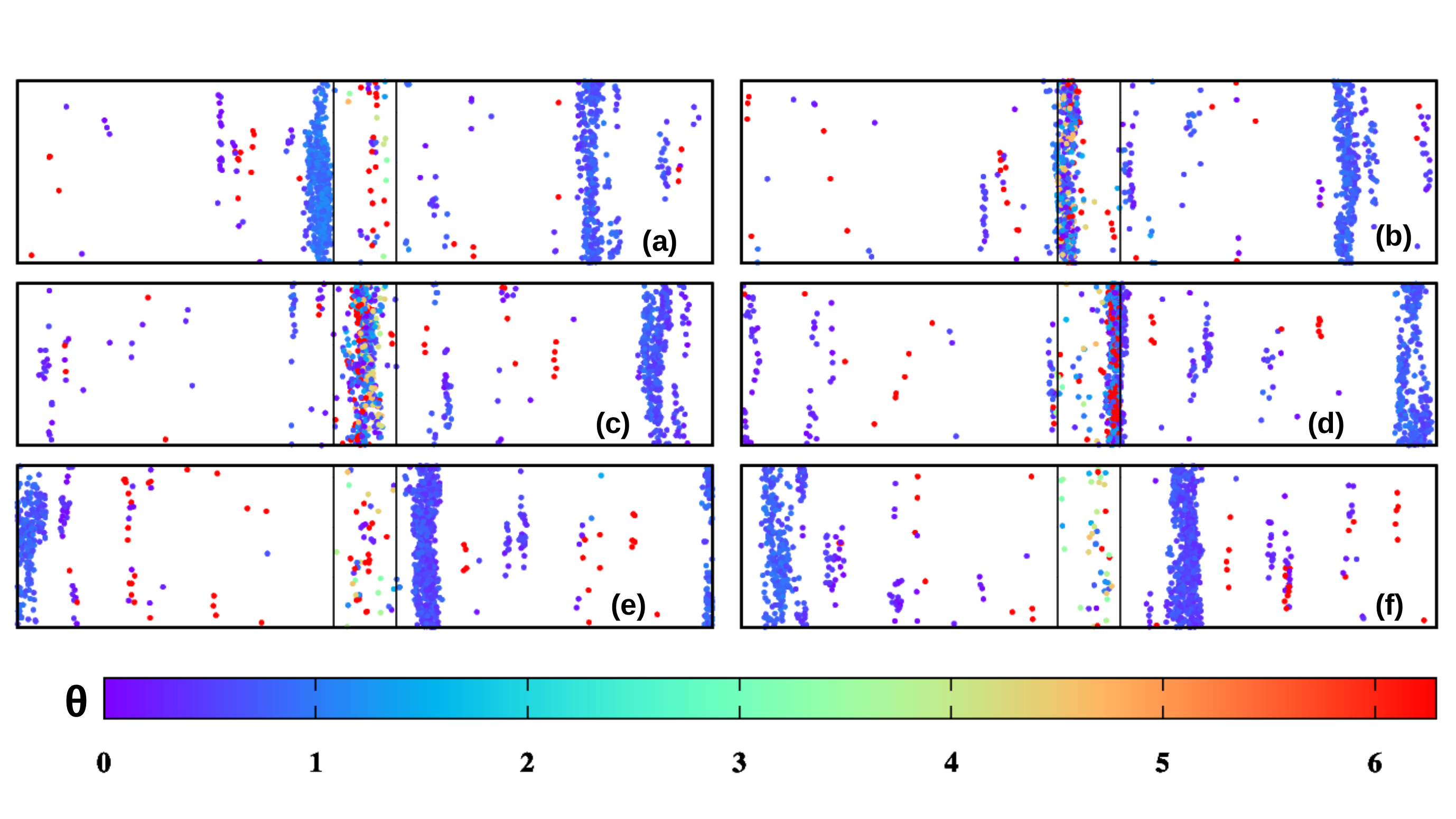}
\caption{(color online) Snapshots of particles with their orientation $\theta$ (shown in colorbar) inside the junction at times $t$ = 2101, 2125, 2150, 2175, 2195 and 2220 from (a-f) 
respectively for system WP ($h_0 = 6\%$). 
There is no clear reflection of the particles from the interface. Other details of the system is the same as in fig.\ref{fig:fig3}.}
\label{fig:fig5}
\end{figure*}
 Fig.\ref{fig:fig4}(b)-(c) show the plot the snapshots for different junction widths $d$ $=2$ and $18$. For lower width of junction flock does not experience any hurdle and passes coherently with different bands, leads to higher values of $\Psi$. Moreover, for higher values of junction width, most of the SPPs trapped into the junction with random directions hence decreases in the value of $\Psi$. This is very different from what we observed for system WOP as shown in fig.\ref{fig:fig2}(c) and fig.\ref{fig:fig3}, where the boundary  of the interface acts like reflecting walls.
 But for the system WP, due to an easy direction for the moving band of SPPs, they are forced to enter inside the interface. Inside the interface the strength of the external perturbation is not strong enough to help them to pass. But it oppose their orientation in other direction. Hence particles feel a kind of frustration inside  and spend more time in the junction as shown in the fig.\ref{fig:fig4}(c) and leads to small order parameter for large junction width $d$. In fig.\ref{fig:fig5}(a-f) we show the interaction of a band moving towards the interface at different times. We found that a moving band experiences some fluctuation from the interface, but density inside the junction is large as 
 shown in fig.\ref{fig:fig5}(c-d), and finally the particles get out of the junction from the other side without getting a clear reflection as found for the case WOP (as shown in fig.\ref{fig:fig5}(e-f)). The details of interaction of  a band with interface for system WP is shown in SM2.
Further we study the properties of the flock along the interface in system without perturbation.
\begin{figure}[ht]
\centering
\includegraphics[width=1.0 \linewidth]{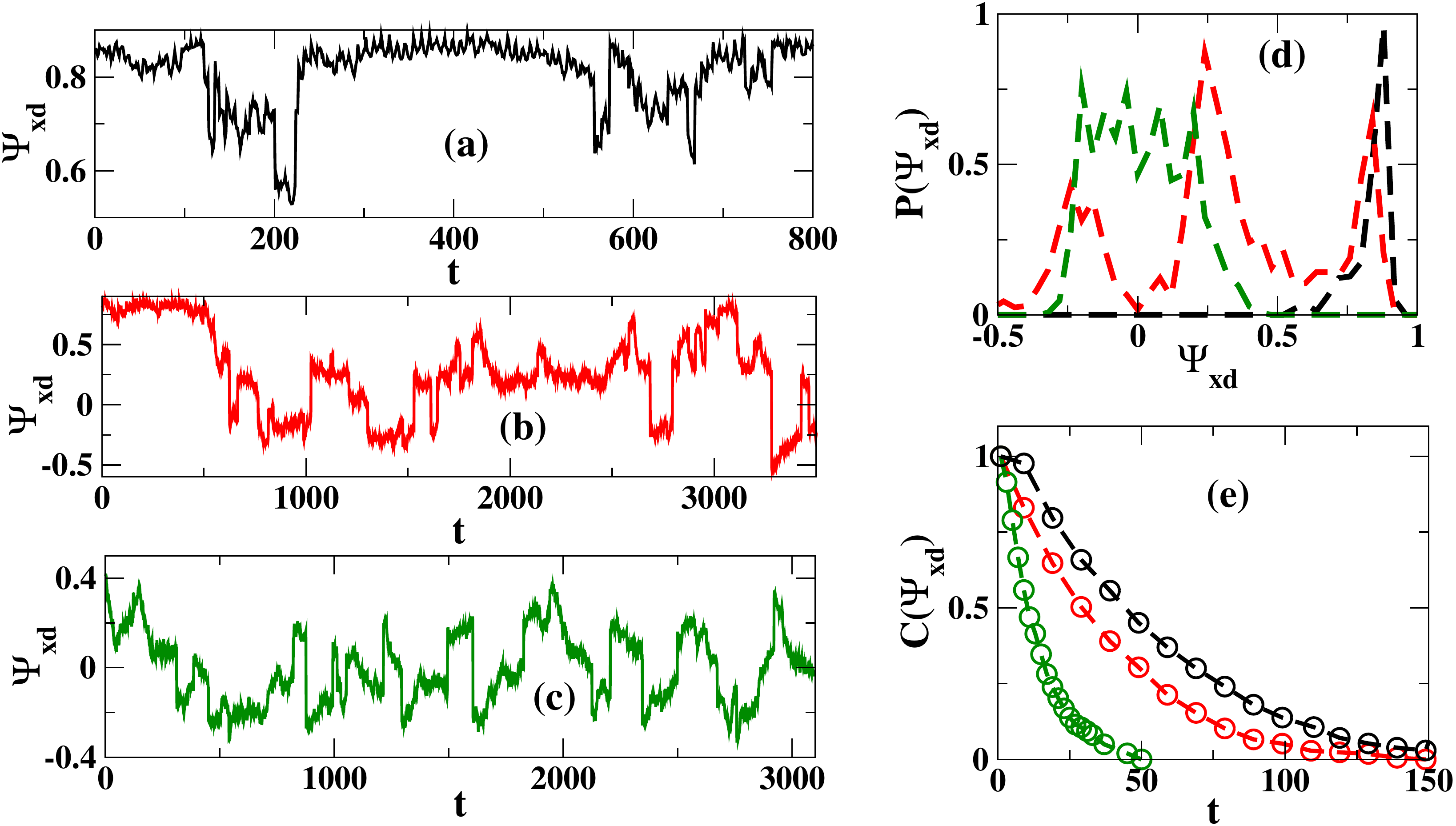}
\caption{(color online) Plot (a)-(c) show the time variation of event current $\Psi_{xd}$ along the long axis with increasing width $d$. (d) event current distribution $P(\Psi_{xd})$. (e) Shows the x-orientation current auto correlation for three junction width $d$. Black, red and green colors show the results for junction width $d$ = 4, 8 and 12 respectively.  $L\times{W}=200\times5$ and $\rho=1.0$.}
\label{fig:fig6}
\end{figure}
\begin{figure}[ht]
\centering
\includegraphics[width=1.0 \linewidth]{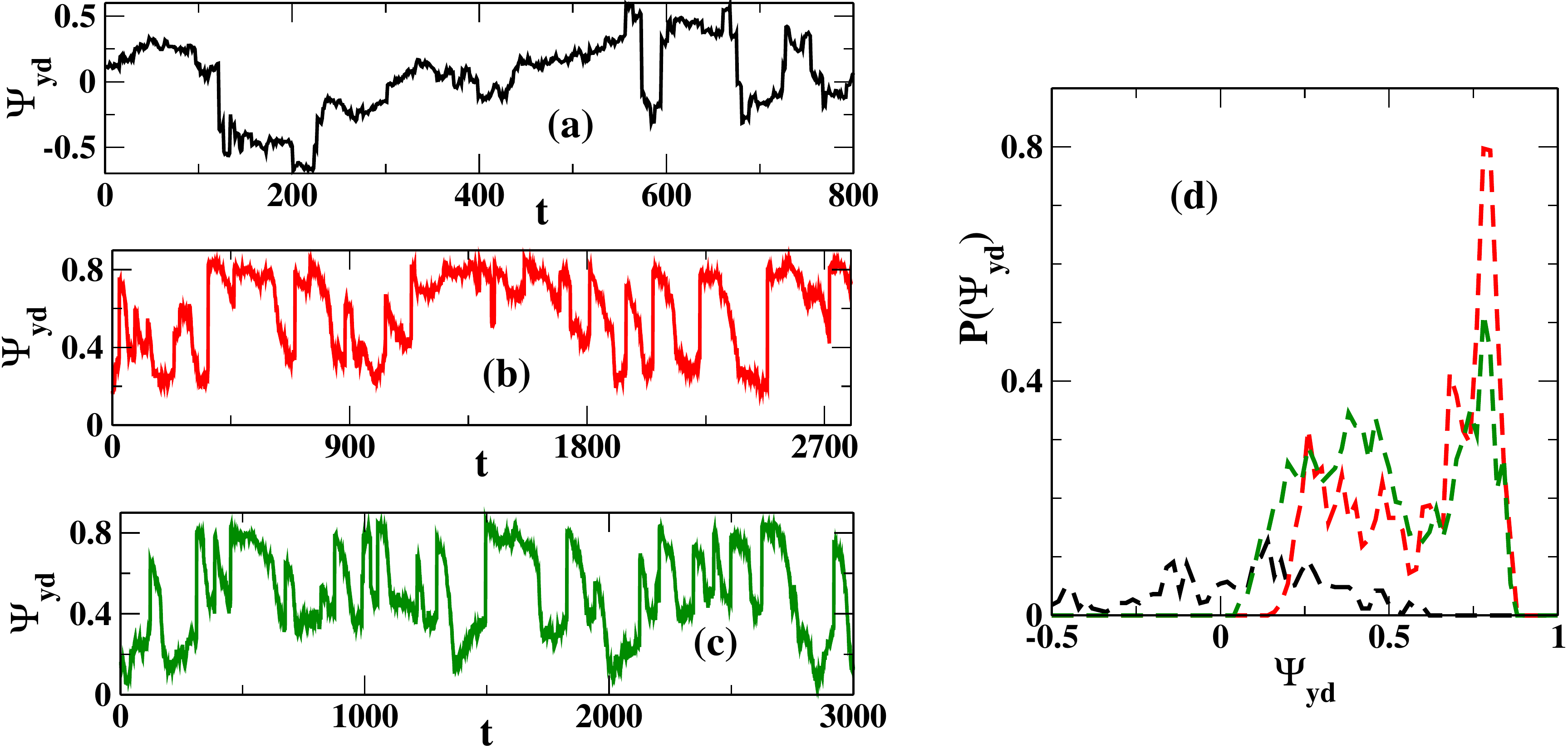}
\caption{(color online) Plot (a)-(c) show the time variation of orientation event current $\Psi_{yd}$ along the short axis with increasing width d. (d) Orientation event current distribution $P(\Psi_{yd})$. Black, red and green colors show the junction width $d$ =4, 8 and 12 respectively.  $L\times{W}=200\times5$ and $\rho=1.0$.}
\label{fig:fig7}
\end{figure}
\subsection{Current inside the junction\label{Properties of flock at the junction}}
{\bf \em Junction current:-} In this section, we discuss the junction current within the junction along the long $x$-axis as well as  $y$-axis 
with the variation of the junction width $d$. 
The junction current is calculated when at least 25\% particles of the whole system are within the junction, and we named this current as event current. The event current in the junction along $x$ and $y$-directions is defined by $\Psi_{xd}(t)=\frac{1}{n}\sum^n_{i}v_{xi}$, $\Psi_{yd}=\frac{1}{n}\sum^n_{i}v_{yi}$ and finally total event current $\Psi_d(t)=\frac{1}{n}\vert{\sum^n_{i}}n_i(t)\vert$.
where $v_{xi},v_{yi}$, and $n$ represent the components of velocity vector along the long and short axis, and the total number of particles within the junction respectively. In the fig.\ref{fig:fig6}(a)-(c) We show the time series of $\Psi_{xd}$ for different values of junction width $d$. We observe, with increased $d$, the amplitude of $\Psi_{xd}$ decreases and also positive and negative current changes in a periodic fashion with the decreased period. Here current is carried by the particles along $+ve$ and $-ve$ $x$-direction, we call them positive and negative currents respectively.  
Further, in fig.\ref{fig:fig6}(d); we show the current probability distribution function $P(\Psi_{xd})$ of $\Psi_{xd}$. It  clearly suggests that with the increase in the width of the junction, there is a clear signature of current reversal. Also, in fig.\ref{fig:fig6}(e) we plot the current-current auto-correlation function $C(\Psi_{xd})=<\Psi_{xd}\cdot \Psi_{xd}>$. Sharper decay of auto-correlation with the increase in the junction width $d$. Hence we find that in the channel the current along the long axis  alternate from $+ve$ to $-ve$ on tuning the width $d$.
Here our claim regarding current reversal phenomena is very interesting properties of the flock at the junction. For small widths $d<8$, coherent flock enters into the junction and crosses without significant deviation. For  
intermediate widths $8\leq d \leq 16$, we observed that once coherent moving SPPs enters into the junction, it faces a randomness inside the junction. Since inside the junction all the directions are equally probable, but flock prefers to move in $+ve$ or $-ve$;  $x$-direction, that lead to the quicker escape from the disorder region. Further SPPs try to come out from the junction and stabilizes back and forth oscillations within the junction. This oscillation along $+ve$ and $-ve$ $x$-direction, we are calling alternating change in the orientation event current. Interestingly this oscillation is more prominent for the intermediate junction widths $8 \leq d \leq 16$. 
For small width of the junction the extend of the moving bands of SPPs is of the order or larger than the width
of the interface and SPPs can easily pass through it with small disturbance hence the $x-$current $\Psi_{xd}$ shows small oscillations with time and no negative current. But as we increase the width of the interface, when the size of the interface is larger than the size of the ordered band, then for some distance the moving band of SPPs is able to penetrate (which is analogous to the penetration depth in solid state) and then experience randomness. Which leads to a fraction of  particles from the moving band reverses its direction of motion and we experience a negative current and hence negative $\Psi_{xd}$ as shown in fig.\ref{fig:fig6}(b-c). This leads to the phenomena of current reversal inside 
the junction. 
For  widths $d >20$, moving SPPs experience more and more reflection and  unable to 
enter inside the junction and gets reflected from the wall itself and hence we have weak junction current. Due to that
the magnitude of the junction current $\Psi_{xd}$ decreases with increasing $d$. 
Furthermore, in fig.\ref{fig:fig7}(a)-(c), we show the junction current $\Psi_{yd}$ and current PDF  $P(\Psi_{yd})$ along the small axis with respect to junction width $d$. We observe that there is no current reversal with the increase in $d$. However, for higher $d$, there is a periodicity which is further confirmed by the current distribution $P(\Psi_{yd})$ is shown shown in fig.\ref{fig:fig7}(d). 
\begin{figure}[ht]
\centering
\includegraphics[width=1.0 \linewidth]{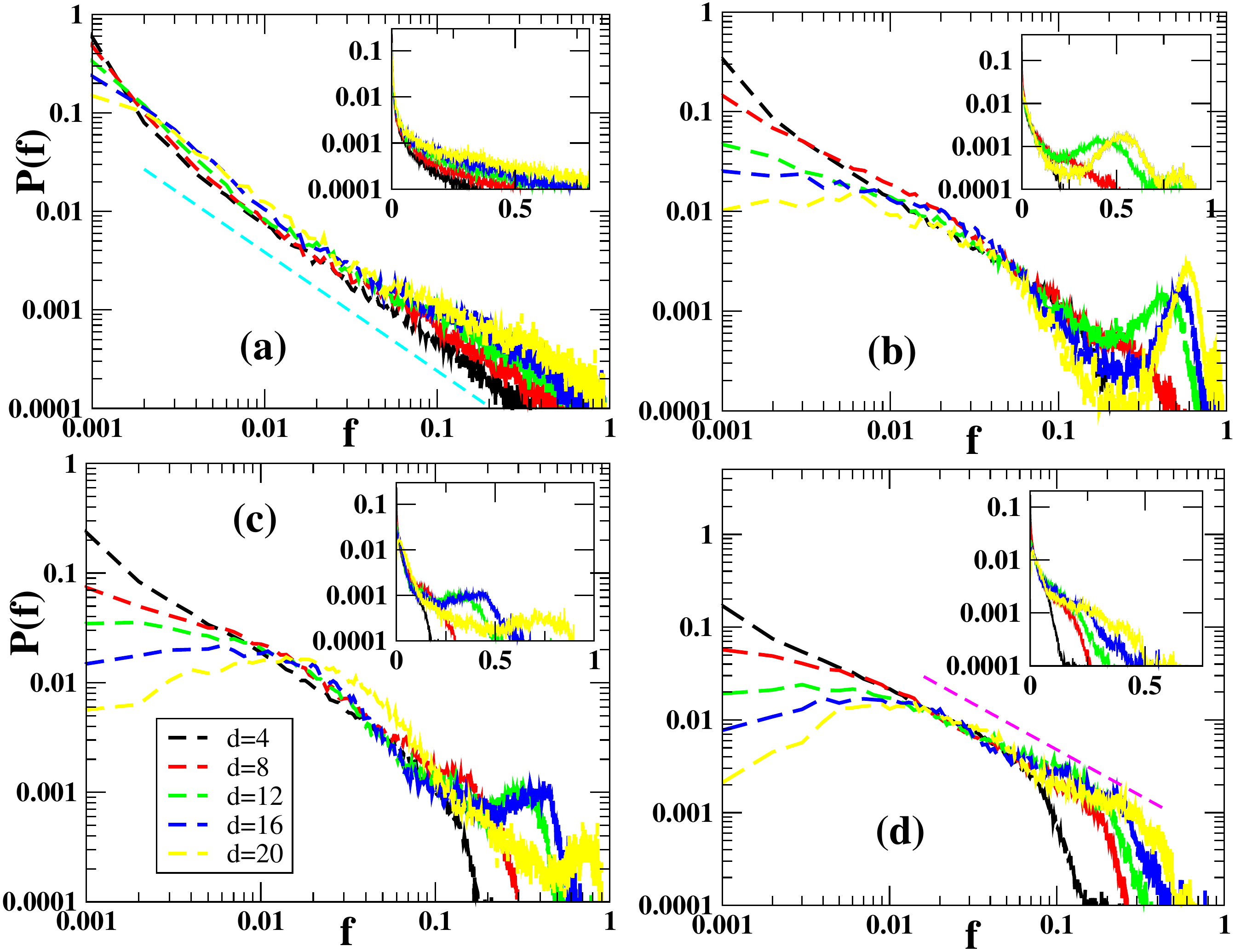}
\caption{(color online) Probability distribution, $P(f)$, inside the junction: Plot (a) is for WOP case and plots (b),(c),(d) are for WP case with perturbation strength ($h_0$) 2$\%$, 4$\%$ and 6$\%$ respectively on $\log$ $\log$ scale. Colors 'black','red','green','blue' and 'yellow' corresponds to $d$ = 4, 8, 12, 16 and 20 respectively. Insets shows the plot of $P(f)$ vs $f$ in $\log-y$ scale. In plot (a) the  dashed line (cyan) is the power law   with exponent 1.25 and in plot (d) dashed line (magenta) is power law with exponent 1. $L \times W = 200 \times 5$ and $\rho =1.0$.
}
\label{fig:fig8}
\end{figure}

\subsection{Fraction of particles inside the junction} \label{Density profile and different time scales within junction}
 Till now, we have discussed about the orientation of moving SPPs inside the junction. We also find interesting results for  
the fraction of particles inside the junction $f(t) = \frac{\mathcal{N}(t)}{N}$. Here $\mathcal{N}(t)$ is the number of particles inside the junction at time $t$. We compared the 
probability distribution function (PDF) of $f(t)$ for the systems WOP and WP. The PDF, $P(f)$ is obtained by calculating the normalised distribution of fraction of particles 
inside the junction and then PDF is averaged different independent realisations.

In fig.\ref{fig:fig8}(a-d) we show the plot of $P(f)$ vs. $f$ for different junction 
widths $(d=4 - 20)$ for the WOP and WP for three different strengths of perturbations $2\%$, $4\%$ and $6\%$ respectively. For all
the cases the  tail of the distribution extends on increasing width of the 
interface $d$, due to the increased area of the interface and hence more number of particles 
inside the junction.
For all junction widths the PDF is power law with slope $f^{-1.25}$ for system WOP (as shown by $\log-\log$ plot in the main plot of fig.\ref{fig:fig8}(a)). It suggests probability of finding 
all possible fractions of particles for system WOP. The inset fig.\ref{fig:fig8}(a) shows the same plot on $\log-y$ scale, to compare that the distribution is clearly power law, and not an exponential. For the system WP,  the distribution clearly has deviation from power law and shows a clear peak for 
 $f$ close to $1$ as shown in fig.\ref{fig:fig8}(b-d). The height of peak increases on increasing width $d$. For larger width and small perturbation $2\%$ (fig.\ref{fig:fig8}(b)) a macroscopic fraction of particles spend time inside the junction. The appearance of peak close to $f\simeq 1$ is visible more clearly in the inset plot of fig.\ref{fig:fig8}(b), which is shown on $\log-y$ scale. As we increase the perturbation further the peak at larger $f$ starts to weaken and distribution get flattens for intermediate $f's$. It starts to appear more exponential in nature for larger perturbations $\ge 4\%$ as shown in the insets of fig.\ref{fig:fig8}(c-d), which are drawn on $\log-y$ scale. The exponential nature of the tail of the distribution represents a critical fraction of particles inside the junction. For $6\%$ perturbation (fig.\ref{fig:fig8}(d) (main figure)), the distribution shows a power law decay  with
power $f^{-1}$ for moderate $f's$. It suggests the moderate fraction of particles inside the junction.\\
 Here we summarise the behaviour of density of particles inside the junction. Adding a finite  perturbation ease the SPPs to get in the interface. For weak perturbation although perturbation is enough for SPPs to enter the junction, but not sufficient for them to overcome the randomness present there. This lead to accumulation of particles inside the junction for some time. Hence a macroscopic fraction of particles spend time inside the junction and $P(f)$ shows a peak at $f \simeq 1$. As we increase the perturbation it lead to quicker entry of SPPs to the junction but now perturbation is comparable to the randomness present inside and they experience frustration inside the junction. This is visible by flattened power law feature of $P(f) \simeq f^{-1}$ for higher perturbations.\\

 Till now, we have discussed the results for the same type of SPPs in the system. In the next section \ref{secsort} we show a one good application of such geometry, where junction can be used for sorting of particles.

 \section{Junction as a particle sorter\label{secsort}}
In this section we  propose that such geometry of system for the case WOP: can also be used for the sorting two type of particles.
As described earlier, for low and intermediate junction widths the particles travel in bands even after passing through the junction.  Because of the periodic boundary condition, with time the band passes through the junction multiple times and after some time the particles separate into multiple narrow bands which  have some dynamics in the transverse direction as well. This motivated us to think:  what will happen if we place a mixture of two different types of particles in the system ?\\
To investigate that we considered a mixture of two different type of particles (1 $\&$ 2) distributed with random orientation and position in the system.
 
Two type of particles differ in their response to the noise: angle of the random vector noise $\phi \in (-0.9\pi, +1.1\pi)$ for one type and   $\phi \in (-1.1\pi,+0.9\pi)$  for the other. Hence one type particles have noise with mean $0.1 \pi$ and the noise has mean  mean $-0.1 \pi$ for the second types. Hence through noise, we have introduced a random clockwise and anticlockwise chirality for the two type of particles. Also the strength of alignment interaction is much stronger for particles of it's own type compared to the other type ($1.0$ and $0.5$ respectively). Results showed that after some time two different types of particles form separate bands (data not shown). 
To characterize this, we define the phase separation order parameter for both types ($P_{1,2}$) as:
\begin{equation}
 P_k= \frac{1}{N_k}\sum_{i=1}^{N_k} \bigg| \dfrac{N_s(i) - N_d(i)}{N_s(i) + N_d(i)} \bigg|
\end{equation}
where,$k=1, 2$ is the index used to denote two different types with $N_1, N_2$ being total number of particles of respective types. For $k^{th}$ type, $N_s(i), N_d(i)$ respectively denote number of particles of similar and dissimilar type inside the interaction radius of $i^{th}$ particle.  

\begin{figure}[ht]
\centering
\includegraphics[width=1.0 \linewidth]{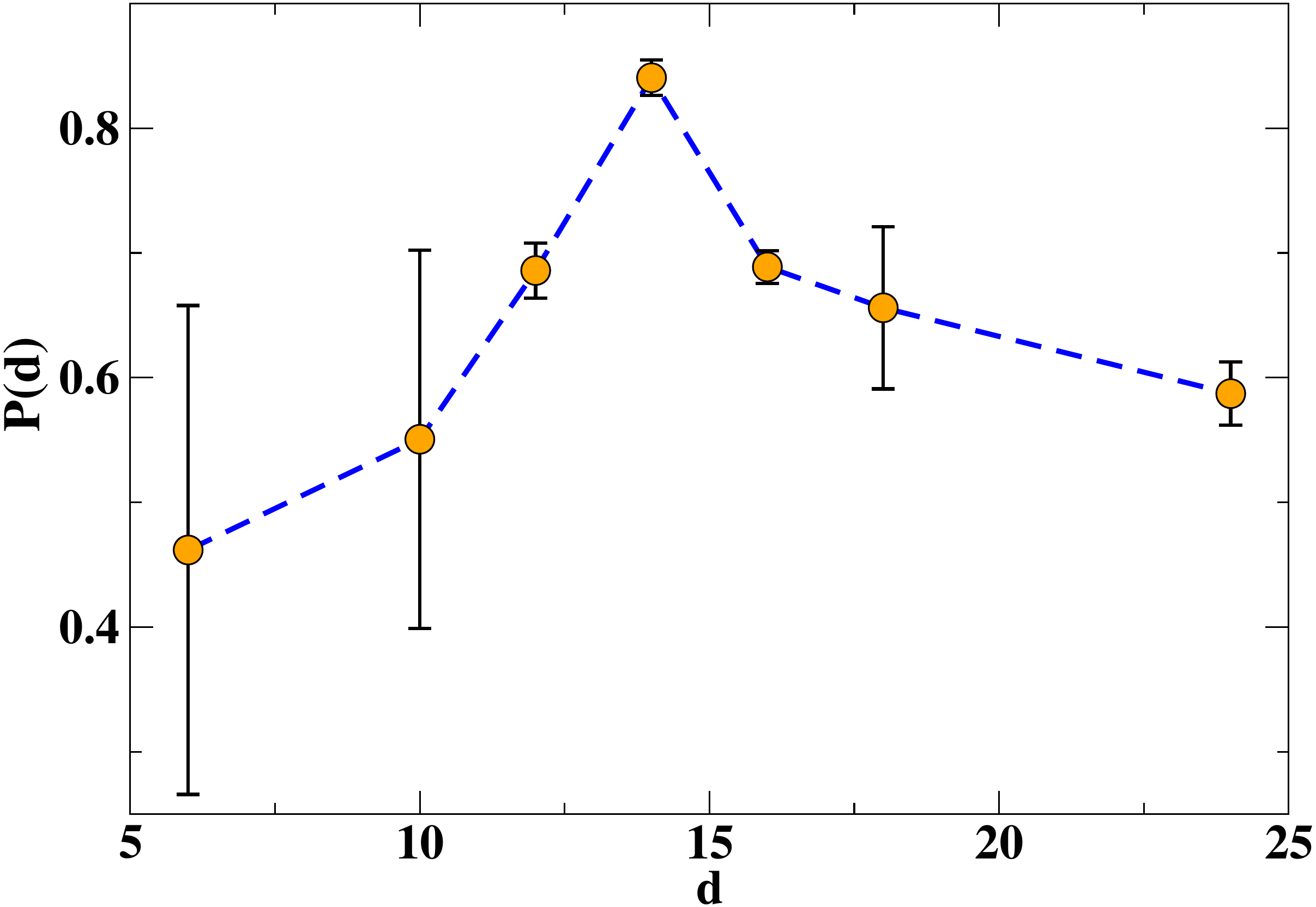}
\caption{(color online) Plot of phase separation order parameter $P(d)$ vs. interface width d. $L \times W = 200 \times 5,  
\rho = 1.0 $}
\label{fig:fig9}
\end{figure}

$P_{k=1,2}$ has value close to $1$ if particles of $k^{th}$ type are separated and is close to $0$ if they are mixed.
We observed that for $d \in (10,15)$ two different types of particles form  bands separated from each other on two sides of the interface and mainly moving along the $y-$ direction.  We  checked the system for two other choices of random noise: $\phi \in (-0.8\pi,+1.2\pi)$ and $\phi \in (-1.2\pi,+0.8\pi)$ for two types and find the same results. In fig.\ref{fig:fig9} we show the plot of $P(d) = <P_{1,2}>$ as a function of junction width $d$.  $< .. >$, stands for average over two type of particles. $P(d)$ shows a peak for $d = 14$ and then decreases on increasing and decreasing $d$ from it. Hence for  intermediate values of interface width $d \simeq 14$, the two types of particles are maximally separated from each other.  Remember this range of $d$ where we find the maximum separation is the same where we find a transition to switching current. \\

\section{Summary and discussion \label{summary}} 
We have studied the properties of collection of  polar self-propelled particles moving on a two dimensional rectangular channel along an order-disorder interface with periodic boundary condition in both directions.  The interaction among the particles is taken as Vicsek type viz; particles move with constant speed and interact through short range alignment interaction. Inside the junction or disorder region, particles experience a high noise disorder state, and outside
they are in the ordered state. The width of the junction is adjusted by the junction width $d$. The model is motivated by the Josephson junction, an analogous equilibrium
system in solid state \cite{JJ1}. We studied the system for the two cases: (i) system WOP, where we do not impose any easy direction for moving SPPs and (ii) system WP where a small external biased direction of motion along the long axis of the channel is introduced. Interestingly, flock experience more disturbance at wider junction width in the system WP in comparison to the system WOP. On increasing width of the junction, the system WOP shows a very small change in the global orientation of particles, whereas for the system WP, it shows transition from macroscopic ordered to disordered state. At the junction, we have found the current reversal for a range of intermediate  widths of
the junction. Such current reversal is due to the reflection of particles from the walls of the interface for intermediate junction widths. 

Further, we also modeled a binary system of two-types of the particle in the system and find that the two type of particles show macroscopic phase separation for the same intermediate width of the interface. Hence such geometry can also be used for the sorting of different types of particles. \\

To the best of our knowledge this kind of study for SPPs at interface has not been explored yet. Although similar setups have been explored in experiments and theory for magnetic devices showing interesting properties \cite{Gurzhi2003,Schmidt2001}. A detail comparison of our results obtained here with these studies is our future work. 
We believe that the results presented here can be tested in experiments by designing such system.  Our study also provide  new scopes in active matter systems where particles experiences different enviournment along their move \\ 
The results presented here can be useful to understand the manufacturing of variety of practical devices using biological agents:  mechanical circuits, switching devices, geophysical sensors, etc.  \\

\section{Acknowledgement}
J.P, P.S.M, V.S. and S. M., thanks PARAM Shivay for computational facility under the National Supercomputing Mission, Government of India at the Indian Institute of Technology, Varanasi and the computational facility at I.I.T. (BHU) Varanasi. P.S.M. thanks UGC for research fellowship. V.S. thanks DST INSPIRE (INDIA) for the research fellowship. S.M. thanks DST, SERB (INDIA), Project No.: CRG/2021/006945, MTR/2021/000438  for financial support.


\begin{thebibliography}{58}
\bibitem [{our()}]{ourdata}%
  \BibitemOpen
  \href@noop {} {\bibinfo  {journal} {For our current system parameters with
  $\rho=1.0$, $v_0=0.5$, $L \times W = 200 \times 5$, with no interface and
  $\eta=0.3$: the typical width of band we found is $\simeq (10)$.}\
  }\BibitemShut {NoStop}%
\end{thebibliography}
\end{document}